\documentclass[fleqn,twoside]{article}
\usepackage[headings]{espcrc2}

\readRCS
$Id: espcrc2.tex,v 1.2 2004/02/24 11:22:11 spepping Exp $
\usepackage{graphicx}
\usepackage[figuresright]{rotating}

\newcommand{\AmS}{{\protect\the\textfont2
  A\kern-.1667em\lower.5ex\hbox{M}\kern-.125emS}}

\title{High-$p_T$ photon processes and the photon structure - results from HERA jet and
  prompt photon (photo)production}

\author{T. Sch\"orner-Sadenius\address[IEXPPH]{Hamburg University, Institute
    for Experimental Physics,\\Luruper Chaussee 149, 22761 Hamburg, Germany}%
        \thanks{Talk given on behalf of the H1 and ZEUS collaborations at the Workshop on `High energy photon collisions at
          the LHC', CERN, Geneva, April 2008.}}
       
\runtitle{High-$p_T$ photon results from HERA}
\runauthor{T. Sch\"orner-Sadenius}

\begin{document}

\begin{abstract}
Many important QCD tests with jets and prompt photons have been performed 
with the experiments H1 and ZEUS at the HERA ep collider. This contribution
focuses on results from jet and prompt photon photoproduction. 
In particular, the concept of resolved photon interactions and various jet cross
sections and their sensitivity to the photon (and proton) PDFs will be
discussed. In addition, recent results from prompt photon production will be
shown. Finally results on multi-parton interactions and the underlying
event will be presented. 
\vspace{1pc}
\end{abstract}

\maketitle

\section{INTRODUCTION}

HERA, the world's only ep collider, stopped operation in June 2007 after 15
years of successful operation. Until that day, the experiments
H1 and ZEUS had each accumulated about 0.5~${\rm fb^{-1}}$ of integrated
luminosity. A major upgrade of machine and detectors in the years 2001 to 2003
(between the so-called HERA-I and HERA-II data taking periods)
proved very fruitful and led to an increase in luminosity of almost a factor
5. 
In the final years of data taking, the proton and  
electron/positron beam energies were 920~GeV and 27.5~GeV, respectively. 

\begin{figure}[t]
\begin{center}
\includegraphics*[width=7pc]{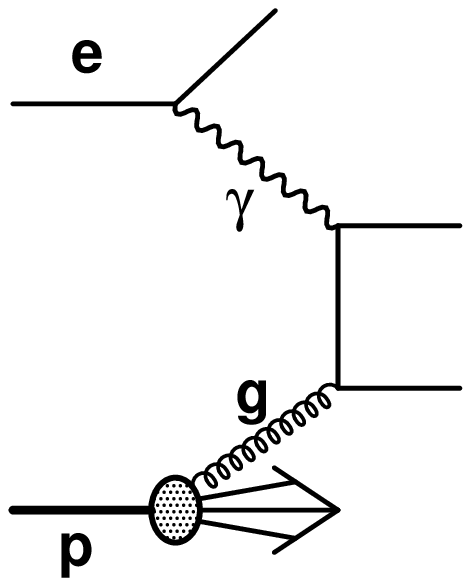}
\includegraphics*[width=10pc]{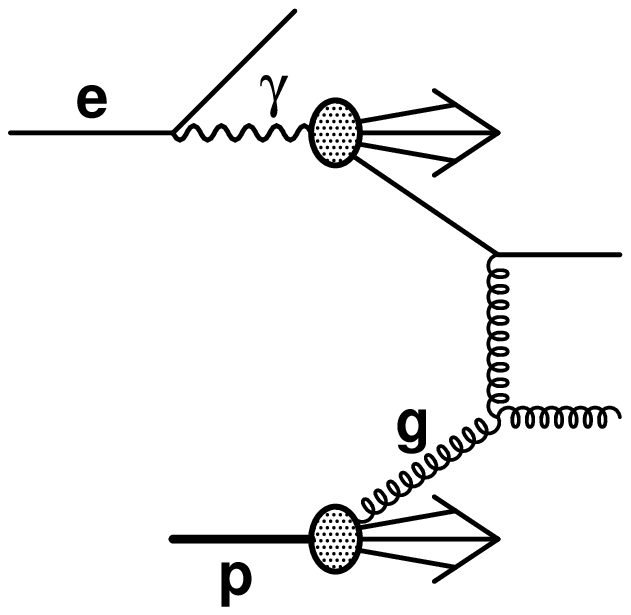}
\caption{Feynman diagrams of direct and resolved dijet photoproduction at LO.}
\label{fig:feynman}
\end{center}
\end{figure}

In photoproduction at HERA, a quasi-real photon emitted from the incoming
electron collides with a parton from the incoming proton. In such events,
hadronic jets and also prompt (meaning: radiated by one of the outgoing quarks)
photons can be produced. The photoproduction
of hadronic jets can be classified into two types of processes in
leading-order (LO) QCD: In direct processes, the entire photon and its
momentum participate in the
hard scatter (left side of Fig.~\ref{fig:feynman}). Resolved processes involve
a photon acting as a source of quarks and gluons, with only a photon momentum
fraction $x_{\gamma}$ participating in the hard scatter (right side of
Fig.~\ref{fig:feynman}). It is due to resolved events that HERA data
might be useful for constraining the photon PDFs further. 

In a more general perspective, the large statistics of jet and prompt photon 
events in photoproduction allows detailed tests of perturbative QCD, with the
transverse 
energy of jets or photons, $E_T$, serving as a hard scale in the QCD
predictions. The concepts
of factorization, of the perturbative expansion of the cross section and of
PDF universality can be tested. In addition, the strong coupling constant
$\alpha_S$ can be extracted from jet photoproduction data. A further issue are
multi-parton interactions and the underlying event which arise in resolved
photoproduction due to the hadronic structure of the photon which makes
photon-proton collisions similar, in some respects, to hadron-hadron
collisions. 

In this contribution, some recent results on jet and prompt photon 
photoproduction at
HERA are discussed. In addition, some results on multi-parton interactions and the
underlying event are reviewed. It should be pointed out that most of
these 
results use data only from the HERA-I data taking period, such that
an improvement in 
statistical precision is to be expected by making use of all available data.

\begin{figure}[t]
\begin{center}
\includegraphics*[width=15pc]{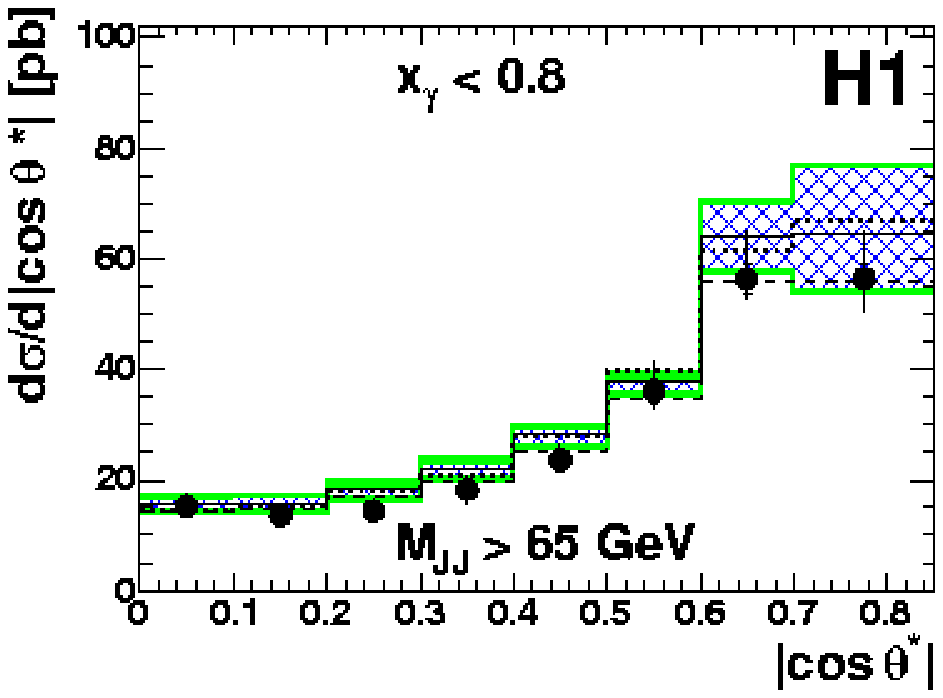}
\includegraphics*[width=15pc]{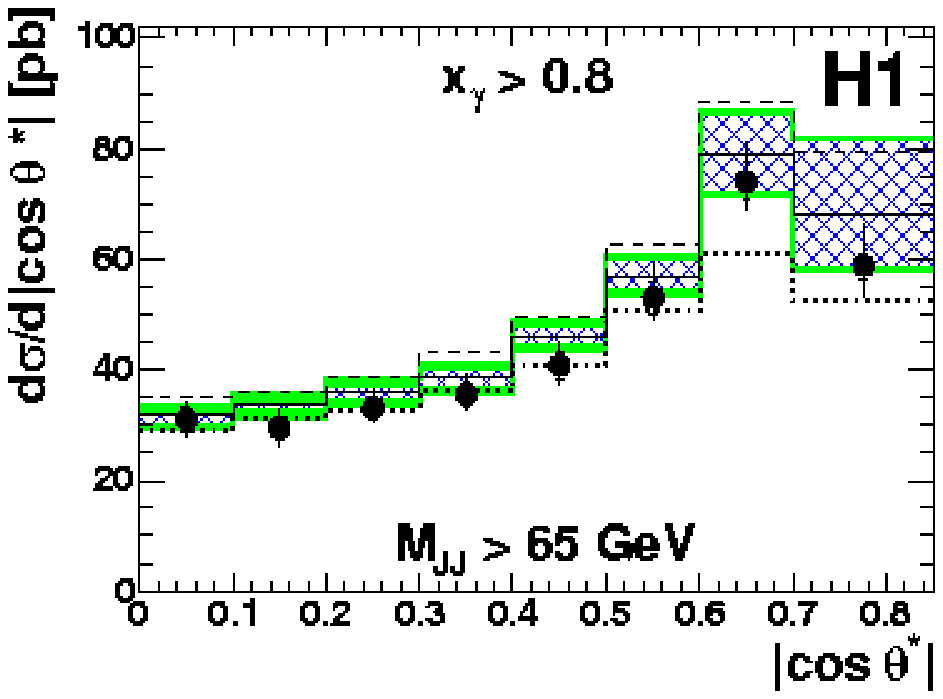}
\caption{Photoproduction dijet cross section as function of CMS scattering
  angle, $\cos\theta^{*}$, for a direct- and a resolved-enriched
  sample~\cite{h1:desy06020}.}
\label{fig:h1cosgam}
\end{center}
\end{figure}

\section{THE CONCEPT OF THE RESOLVED PHOTON}

Fig.~\ref{fig:feynman} shows Feynman diagrams for direct (left) and resolved
(right) photoproduction of dijets. Statistically, direct events are dominated
by quark propagators whereas resolved events are mostly characterized by gluon
propagators. This difference should lead to a distinctly different angular
behaviour of the final-state jets: Whereas the quark propagator (quarks being
spin-$1/2$ particles) should lead to
a distribution in the cosine of the CMS scattering angle, $\cos\theta^{*}$, like
$\left(1-|\cos\theta^{*}|\right)^{-1}$, in the gluon case a distribution
like $\left(1-|\cos\theta^{*}|\right)^{-2}$ is expected. In other words, the
cross section of the resolved part is expected
to rise more rapidly towards higher $\cos\theta^{*}$ than that of the direct
part. Fig.~\ref{fig:h1cosgam} shows the experimental
evidence~\cite{h1:desy06020}: Shown is the cross section as function of
$\cos\theta^{*}$ for a direct-enriched (left) and a resolved-enriched data
sample (right). It
is obvious that the above predictions are fulfilled, the resolved distribution
rising much more rapidly than the direct one. These distributions thus form
an important test of the concept of the resolved photon (similar results have
been obtained by the ZEUS collaboration~\cite{zeus:desy01220}). 

\begin{figure}[t]
\begin{center}
\includegraphics*[width=17pc]{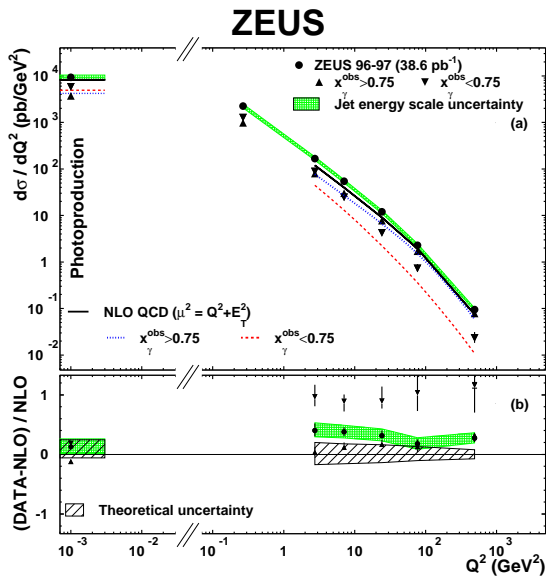}
\caption{Fraction of resolved events as a function of
  $Q^2$~\cite{zeus:desy04053}.}
\label{fig:zeusfractionresolved}
\end{center}
\end{figure}

In the above discussion, the distinction between direct and resolved data
samples has been made. On the theoretical side, this
distinction is meaningful only at LO. On the experimental side, the
distinguishing observable $x_{\gamma}$ is not directly accessible but has to
be reconstructed from the two final-state jets in much the same way as the
proton's momentum fraction $x_p$ entering the hard scattering, 
\begin{eqnarray*}
x_{\gamma} = \frac{E_{T,1}e^{-\eta_1}+E_{T,2}e^{-\eta_2}}{2yE_e}, \\ 
x_{p} = \frac{E_{T,1}e^{+\eta_1}+E_{T,2}e^{+\eta_2}}{2E_p},
\end{eqnarray*}  
with the jet transverse energies and pseudorapidities $E_{T,i}$ and $\eta_i$,
the inelasticity $y$ (characterizing the energy loss of the scattered electron)
and the electron and proton beam energies $E_e$ and $E_p$. Typically, the
resolved regime is defined to comprise values of $x_{\gamma}$ between 0 and 0.75 or
0.8.

However, the phenomenon of the resolved photon is not strictly confined to the
photoproduction regime. Also the virtual photon entering into deep-inelastic
scattering (DIS) events can exhibit a hadronic substructure, leading to a
resolved contribution to DIS. The ZEUS collaboration has evaluated the
fraction of resolved events in both photoproduction and DIS, measuring the
fraction of dijet events with $x_{\gamma}$ below and above 0.75. The results
are shown in Fig.~\ref{fig:zeusfractionresolved} as a function of
the photon virtuality $Q^2$~\cite{zeus:desy04053} and are compared to
next-to-leading order (NLO) QCD
calculations. It is found that even at large $Q^2$ values (highly virtual
photons), there is a significant contribution from resolved events and that for
DIS this component of the data is not correctly described by the QCD
predictions which do not include the resolved photon option. In contrast, the
resolved contribution to photoproduction is well described by
QCD. 

\begin{figure}[t]
\begin{center}
\includegraphics*[width=18pc]{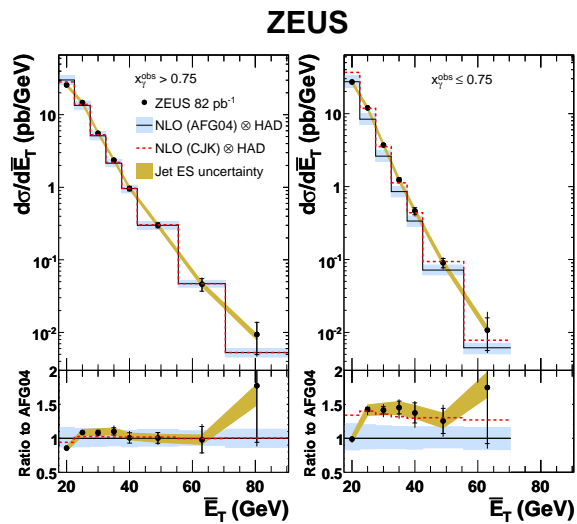}
\caption{Photoproduction dijet cross sections as functions of the mean
  transverse jet energy, $\overline{E}_T$, for a direct- and a resolved
  enriched sample~\cite{zeus:desy07092}.}
\label{fig:hannoetmean}
\end{center}
\end{figure}

\section{JET CROSS SECTIONS IN PHOTOPRODUCTION}

Numerous measurements of inclusive-jet, dijet and multijet cross sections have
been performed by the  HERA experiments. A very recent result is presented in
Fig.~\ref{fig:hannoetmean}~\cite{zeus:desy07092}. Shown is the dijet cross
section as a function of the mean dijet transverse energy, $\overline{E}_T$,
separately for a sample enhanced in direct and a sample enhanced in resolved
events. The data are compared to an NLO QCD prediction using two different
parametrizations of the photon PDFs. 

\begin{figure}[hhhh]
\begin{center}
\includegraphics*[width=17pc]{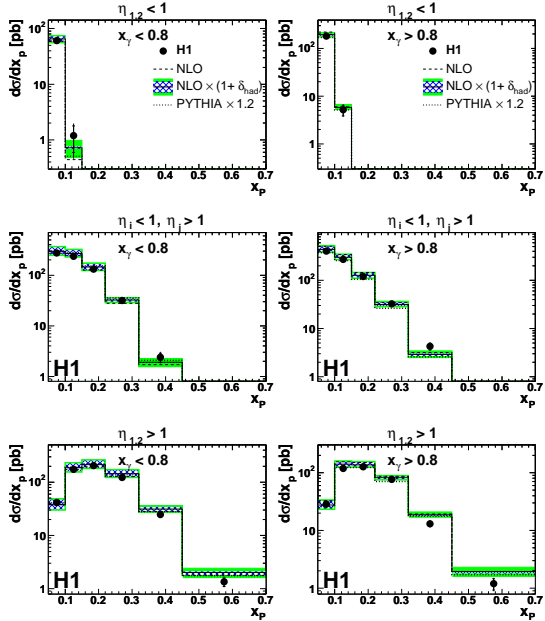}
\caption{Photoproduction dijet cross sections as functions of the momentum
  fraction $x_p$ for a direct- and a resolved
  enriched sample~\cite{h1:desy06020}.}  
\label{fig:desy06020a}
\end{center}
\end{figure}

Especially the data in the direct regime
are very well described by the theory (on the level of 10~$\%$ or better), 
as can be seen in the bottom left part
of the figure which shows the ratio of data over NLO prediction. The
uncertainties here are dominated by the theory uncertainty which is of the order
of 15~$\%$. The situation in the resolved regime is slightly more complicated
and will be discussed in some more detail below.

\begin{figure}[t]
\begin{center}
\includegraphics*[width=18pc]{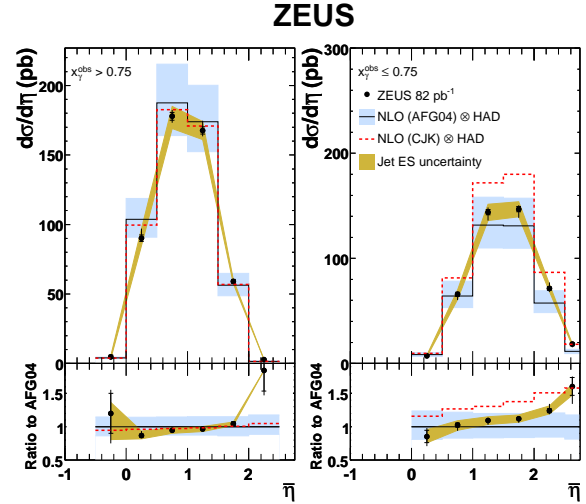}
\caption{Photoproduction dijet cross sections as functions of the mean jet
  pseudorapidity, $\overline{\eta}$, for a direct- and a resolved
  enriched sample~\cite{zeus:desy07092}.}  
\label{fig:hannoeta}
\end{center}
\end{figure}

Many more examples of photoproduction jet cross sections and their successful
description by NLO QCD exist. Fig.~\ref{fig:desy06020a}~\cite{h1:desy06020} 
shows the dijet cross section as a function of
$x_p$ in different regions of $x_{\gamma}$ and of the jet
pseudorapidities. Both the momentum fractions and the pseudorapidity
distributions of the jets are sensitive to the momentum distributions of
partons inside the proton, making these measurements important tests of
QCD. It can be seen that, again, the data are very well described by NLO QCD,
on the level of 10~$\%$, which are well covered by the combined uncertainties.
Only for large $x_p$ values with both jets going forward ($\eta_{1,2} >
1$) some deviations between data and theory occur 
(Fig.~\ref{fig:desy06020a}, bottom right) which might be
explained by the large uncertainties on the proton
PDFs for large momentum fractions.

Fig.~\ref{fig:hannoeta}~\cite{zeus:desy07092} shows, for the same data
sample as in Fig.~\ref{fig:hannoetmean}, the cross section as a function of the
mean jet pseudorapidity, $\overline{\eta}$. The data are again shown
separately for a direct and a resolved sample and are compared to NLO QCD
predictions using different parametrizations of the photon PDFs. For the direct
case, the description of the data by the theory is again excellent. 

The demonstrated good performance of NLO QCD in describing photoproduction
data (especially in the direct regime) gives confidence in the theory, thus
rendering possible the extraction of QCD parameters like the strong coupling
constant $\alpha_S$ or the proton and photon PDFs from the data. One example
for the former is given in \cite{zeus:desy02228} where a value $\alpha_S =
0.1224\pm 0.0001(stat.)^{+0.0022}_{-0.0019}(exp.)\pm^{+0.0054}_{-0.0042}(th.)$
was extracted from an inclusive-jet measurement in photoproduction. The impact
of jet cross sections on the PDFs is discussed in the next section.

\begin{figure}[t]
\begin{center}
\includegraphics*[width=8pc]{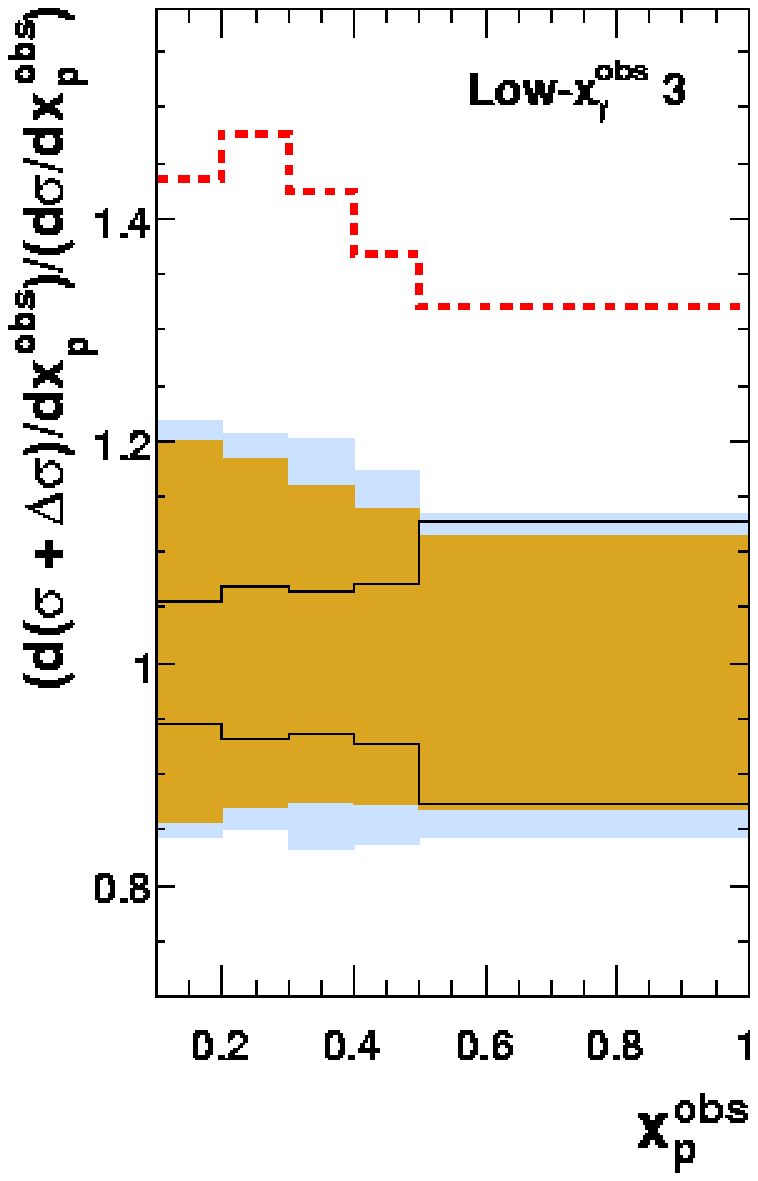}
\includegraphics*[width=8pc]{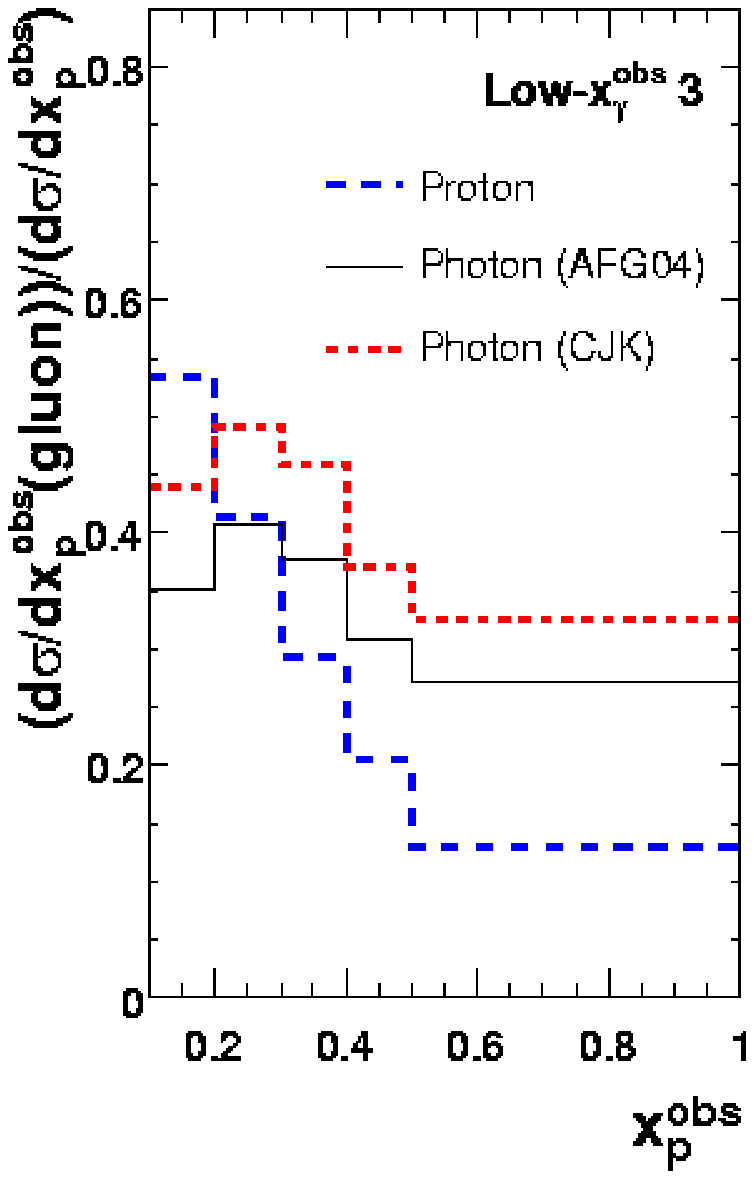}
\caption{Left: Theoretical uncertainties on photoproduction dijet cross
  sections in a special kinematic region~\cite{zeus:desy07092}. 
  Right: Gluon-induced contributions to photoproduction dijet cross
  sections in the same kinematic region.}
\label{fig:hannotheoerror}
\end{center}
\end{figure}

\section{JETS IN PHOTOPRODUCTION AND THE PDFs}

\begin{figure}[t]
\begin{center}
\includegraphics*[width=17pc]{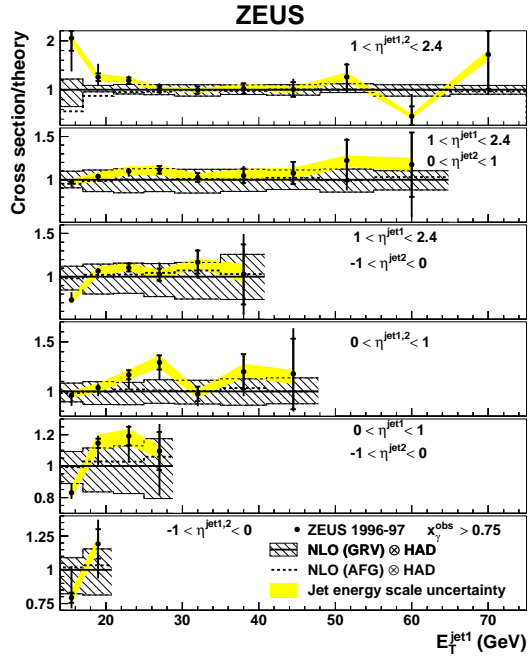}
\caption{Ratio of data over NLO QCD predictions for photoproduction dijet
  events in the direct regime as function of jet transverse energy and
  pseudorapidity, $E_T$ and $\eta$~\cite{zeus:desy01220}.}  
\label{fig:matthewfig5}
\end{center}
\end{figure}

In~\cite{zeus:desy07092} both the theoretical uncertainties on dijet cross
sections and their sensitivity to the gluon density in the photon and the
proton have been investigated in great detail. As is highlighted in
Fig.~\ref{fig:hannotheoerror} (left) for a special choice of kinematics, there
are regions in which the proton PDF uncertainty (indicated as the region
between the two solid lines) is as large as or even larger than the combined
uncertainty from the variation of the renormalization scale and the PDF
uncertainties (indicated as the coloured area). Also the
uncertainty due to the very imprecise knowledge of the photon PDF may be very
large, reaching values of up to 60~$\%$, as is indicated by the
dashed line in the figure which shows the difference in the cross section
prediction between two different photon PDF parametrizations. Dijet data
therefore do have the potential to further constrain both quark and 
gluon densities in the photon and the proton,
as is indicated in Fig.~\ref{fig:hannotheoerror} (right). The figure shows the
fraction of gluon-induced events on the proton side (dark dashed line) and on
the photon side according to two different photon PDF parametrizations (light
dashed and solid lines). The amount of gluon-induced events can be as large as
60~$\%$, depending on the detailed kinematics under consideration.  

The large discrepancies between different photon PDF parametrizations is also
visible in the comparison of NLO QCD predictions with dijet cross sections in
the resolved regime like in Fig.~\ref{fig:hannoetmean} (right) or
Fig.~\ref{fig:hannoeta} (right). For example, in Fig.~\ref{fig:hannoetmean},
the resolved dijet cross section can be approximately described by the NLO
prediction using the CJK photon PDF parametrization, but not by the AFG04
parametrization which is off by up to 40-50~$\%$. This difference highlights
the potential of the data to further constrain the photon PDFs. 

\begin{figure}[t]
\begin{center}
\includegraphics*[width=16pc]{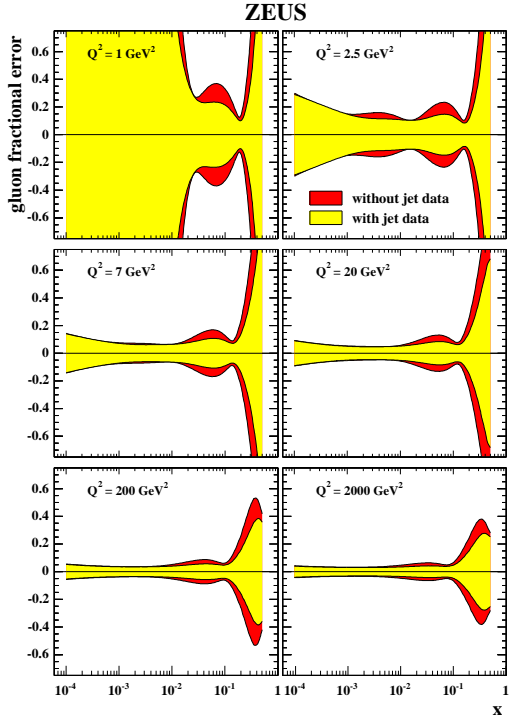}
\caption{Improvement of gluon density by using jet data in NLO QCD fits of the
  PDFs~\cite{zeus:desy05050}.}  
\label{fig:fitpaper}
\end{center}
\end{figure}

A first example of the benefit of jet photoproduction data on determinations
of the proton PDFs is given in Figs.~\ref{fig:matthewfig5}
and~\ref{fig:fitpaper}. Fig.~\ref{fig:matthewfig5} shows, for an older
measurement of photoproduction dijet cross sections, the ratio of the measured
cross sections over the NLO prediction. An overall good description is found,
with data and theory in agreement almost everywhere within the combined
theoretical and experimental uncertainties. These data (together with data
from DIS jet analyses) have been used as additional inputs (besides the usual
inclusive $F_2$ data) to an NLO QCD PDF fit. The success of this fit is
demonstrated in Fig.~\ref{fig:fitpaper} which shows the fractional gluon
density uncertainty as a function of the proton momentum fraction $x$ in
different regions of $Q^2$. The uncertainty without the use of jet data is
given by the dark shaded area, and the result including jet data is given by
the light shaded area. An improvement in the uncertainty of up to
35~$\%$ is clearly visible especially in the region of medium/high $x$
values. 

The aim is to further improve the proton PDFs (and here especially the gluon
density at high values of $x$ as this is especially important for 
LHC physics) by using
more precise or different cross sections from both photoproduction and
DIS. Constraining the photon PDFs will be technically even more demanding, 
partly because of
lack of a consistent PDF error treatment for the photon PDF, partly because
of the increased experimental and theoretical uncertainties for the resolved
regime. 


\section{PROMPT PHOTON PRODUCTION}

The production of prompt photons from the hadronic final state offers an
alternative access to the QCD dynamics in ep scatterings, with different
systematic uncertainties and reduced effects from hadronisation. Prompt photon
production has been measured by ZEUS and H1 in both DIS and photoproduction;
in addition to inclusive measurements of prompt photons, often also photon+jet
cross sections are measured for which there are also 
NLO QCD predictions available. 

\begin{figure}[t]
\begin{center}
\includegraphics*[width=11pc]{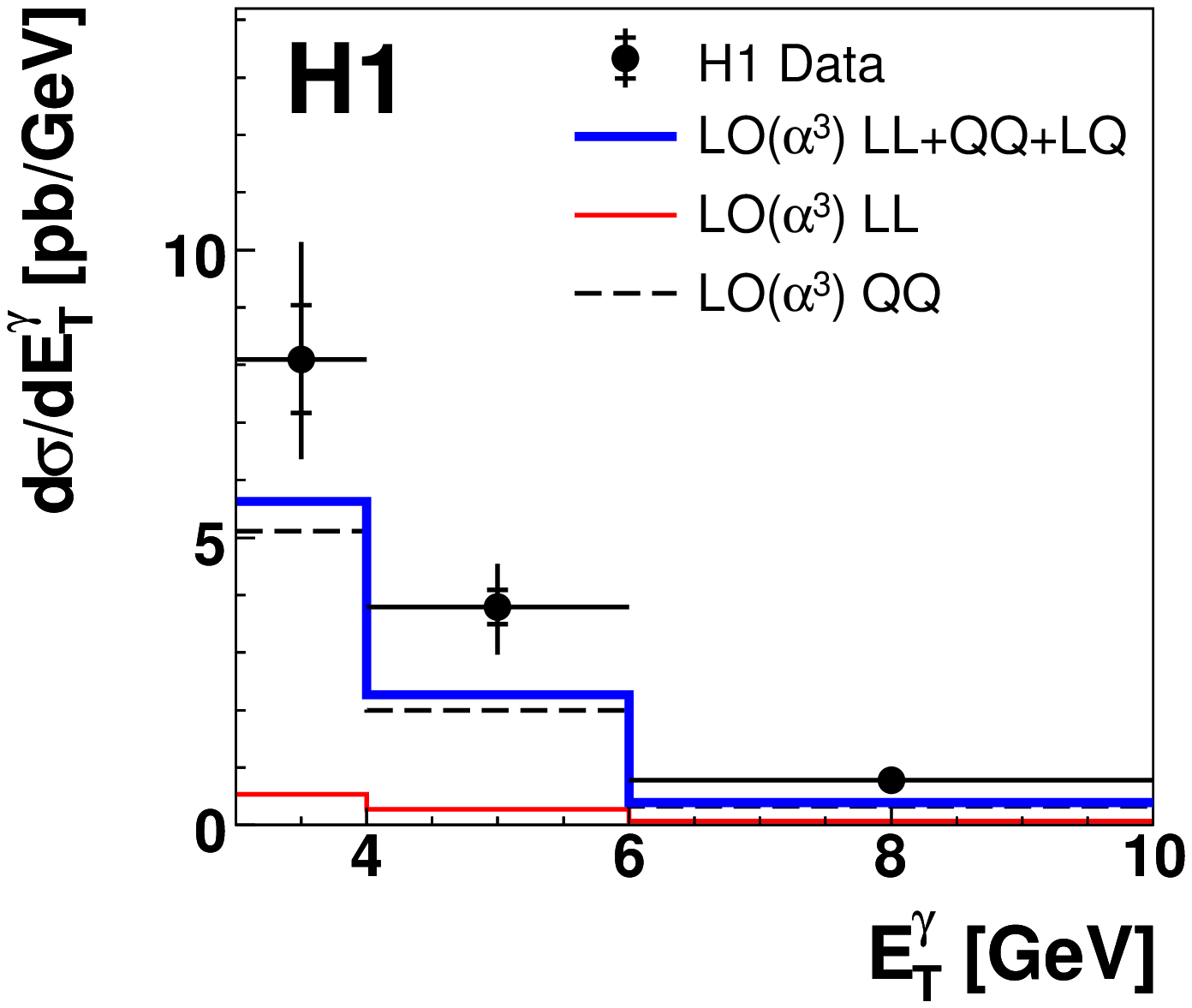}
\includegraphics*[width=11pc]{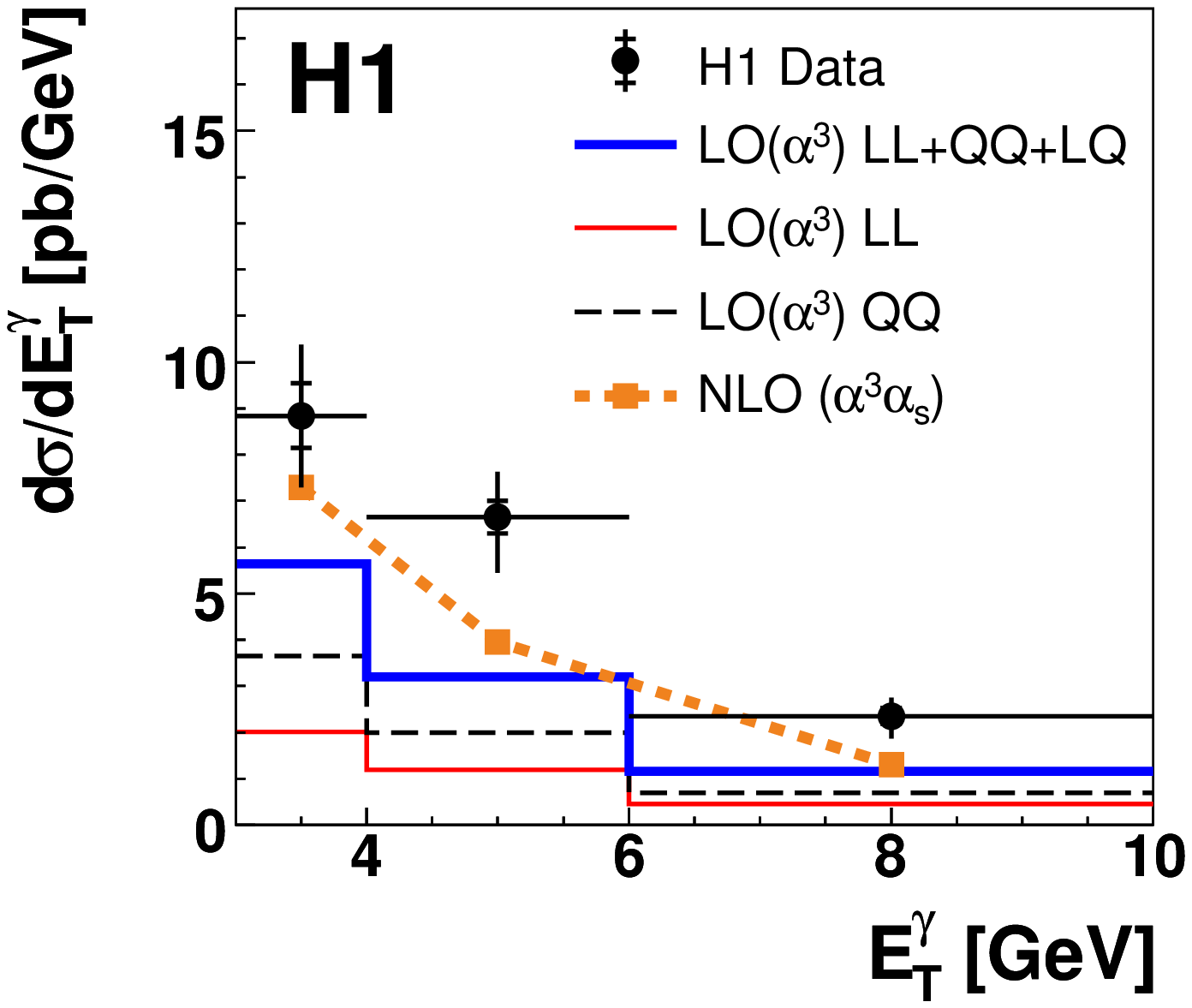}
\caption{Prompt photon production in DIS
  events~\cite{h1:desy07147}. Shown are the cross sections as functions of
  photon transverse energy, $E_T^{\gamma}$, for an inclusive photon and a
  photon+jets sample.}  
\label{fig:h1promptdis6}
\end{center}
\end{figure}

Fig.~\ref{fig:h1promptdis6}~\cite{h1:desy07147} shows the inclusive photon
cross section (left) and the photon+jet cross section (right) in DIS 
as a function of
the photon transverse energy, $E_T^{\gamma}$. It can be seen that all LO
predictions are not able to describe the data. In contrast to that, the NLO
theory curve in the photon+jet sample is compatible with the data. 


Similar results are obtained for the photoproduction
case~\cite{h1:desy04118},~\cite{zeus:desy06125}. Again, NLO calculations are
needed to describe the data, and the additional requirement of a jet in
addition to a prompt photon brings data and theory closer together.


In summary, prompt photons might serve as a valuable place for QCD tests. Currently, however, the
predictions are not yet in a state as to allow (for example) the use of prompt
photon data in NLO QCD fits of the PDFs or for $\alpha_S$ extractions. 

\begin{figure}[t]
\begin{center}
\includegraphics*[width=17pc]{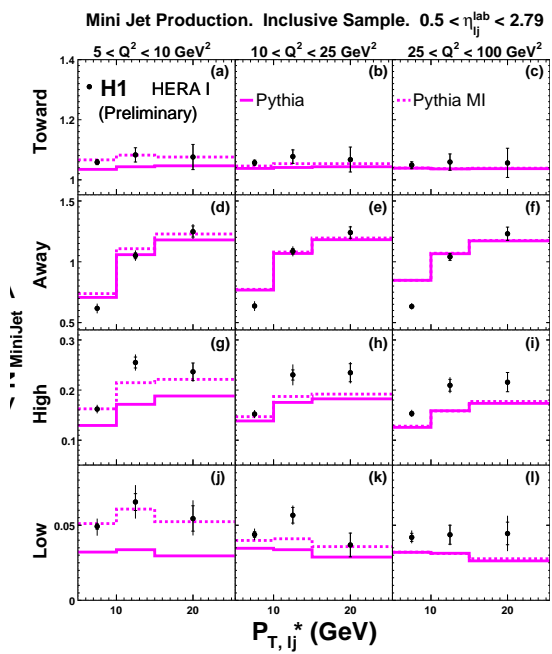}
\caption{Mean number of minijets as function of the leading jet
  $p_T$~\cite{h1:prelim07032}.}  
\label{fig:h1minijets2}
\end{center}
\end{figure}

\section{THE UNDERLYING EVENT AND MULTI-PARTON INTERACTIONS}

As has been pointed out in the introduction, resolved photon-proton
interactions may, in some respects, be regarded as hadron-hadron collisions,
with all the additional features with respect to direct interactions. In
particular, in hadron-hadron collisions it is possible to have multiple
interactions of pairs of partons (`multi-parton interactions', MPI) which may
populate the hadronic final state with additional soft or hard jets or
additional energy flow throughout the detector. This effect may alter the
final state significantly, making it necessary to model it adequately in the
Monte Carlo programs used in the analyses. There exist various MPI model
implementations 
in the standard generators
HERWIG and PYTHIA which can be tested against data or whose parameters can be
adjusted to describe the data. Here, two recent examples of MPI studies at
HERA will be briefly discussed. 

Fig.~\ref{fig:h1minijets2}~\cite{h1:prelim07032} shows the the mean number of
`minijets' (i.e. soft jets with transverse energies above a very low cut of
3~GeV) in events with at least one hard jet. The data are shown as a function
of this leading jet's transverse momentum, $p_{T,lj}^*$, in different regions of
$Q^2$ and different azimuthal-angle regions with respect to the leading jet's
azimuth. The `Towards' and 'Away' regions are supposed to be mostly populated
by the results of the first and hardest parton-parton scattering in the event,
the dijet system coming from this scattering supposed to be separated by about
$\pi$ in azimuth. The `High' and especially the `Low' regions, in contrast,
are supposedly particularly sensitive to MPI effects which in these regions
are not masked out by the harder energy depositions from the leading jet
pair (`High' and `Low' refer to the amount of deposited energy in the two
regions). 
It can be observed that the PYTHIA model with MPI effects switched on (`PYTHIA
MI') is in rather good agreement with the data in almost all regions. In
contrast PYTHIA without MPI modeling (`PYTHIA') fails to describe the data in
the low $Q^2$ `High' and `Low' regions, consistent with the hypothesis of
dominating MPI effects in these regions of phase space. The data thus
clearly indicate the necessity of MPI effects in the models. 

\begin{figure}
\begin{center}
\includegraphics*[width=14pc]{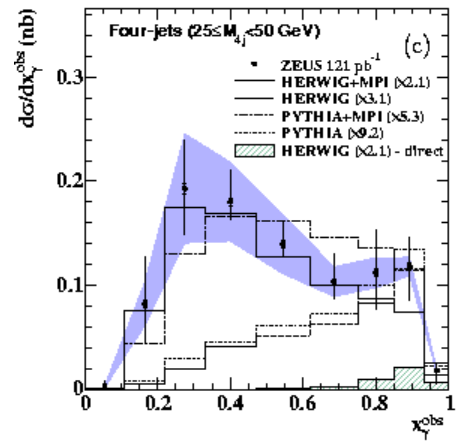}
\caption{Photoproduction fourjet cross section as function of $x_{\gamma}$~\cite{zeus:desy07102}.}  
\label{fig:tim4}
\end{center}
\end{figure}

A similar statement is derived from a recent measurement of three- and fourjet
photoproduction~\cite{zeus:desy07102}. Fig.~\ref{fig:tim4} shows the fourjet
photoproduction cross section as a function of $x_{\gamma}$ and compares it to
PYTHIA and HERWIG predictions with and without the inclusion of MPI
modeling. It becomes clear that only the two predictions including MPI
effects (`HERWIG+MPI' and `PYTHIA+MPI') can reproduce the data, whereas the
models without MPI grossly underestimate the data in the resolved regime (for
$x_{\gamma} <$~0.8). The effect is particularly drastic for low
energy scales and low invariant multijet masses, as in Fig.~\ref{fig:tim4}
where the fourjet mass was between 25 and 50~GeV. Again, the
need for MPI contributions in the models is evident. 


However, the HERA data so far do not have the power to specify more precisely
the mechanism underlying MPI effects or to shed light on the energy evolution of
MPI effects when going (for example) from TEVATRON to LHC centre-of-mass
energies. The models in use so far are rather
crude and will have to be replaced by more realistic models and calculations
which take correctly into account features like multi-parton exchanges between
photon and proton, correlations between these exchanges, etc. 

\section{CONCLUSIONS}

Several aspects of jet and prompt photon production (mainly) in
photoproduction at HERA have been discussed. The power of NLO QCD in
describing jet cross sections (and also prompt photon data) has
been shown, and the potential of the data for further constraining the proton
(and photon) PDFs has been pointed out. In addition, some effects of MPI on
HERA data have been discussed.

\end{document}